\documentclass[aps,showpacs,twocolumn,superscriptaddress]{revtex4}
\usepackage{amssymb}
\usepackage{natbib}
\usepackage{amsmath}
\usepackage{amsthm}
\usepackage{amsfonts}
\usepackage{graphicx}
\usepackage{mathrsfs}
\usepackage{dcolumn}
\usepackage{bm}
\usepackage{color}
\definecolor{rot}{rgb}{0.75,0.05,0.25}
\definecolor{hellgrau}{gray}{0.5}
\definecolor{blau}{rgb}{0,0,0.7}

\definecolor{rot}{rgb}{0.75,0.05,0.25}
\definecolor{hellgrau}{gray}{0.5}
\definecolor{blau}{rgb}{0,0,0.7}

\newtheorem{statement}{Statement}
\newtheorem{theo}{Theorem}

\newtheorem*{hypo}{Ergodic Hypothesis}

\begin{document}

\title{Derivation of Boltzmann Principle}
\author{Michele Campisi}\email{Michele.Campisi@physik.uni-augsburg.de}
\affiliation{Institute of Physics, University of Augsburg, Universit\"atsstrasse 1, D-86153 Augsburg, Germany}
\author{Donald H. Kobe}
\affiliation{Department of Physics, University of North Texas, P.O. Box 311427, Denton, TX 76203-1427, USA}
\date{\today }

\begin{abstract}
We present a derivation of Boltzmann principle $S_{B}=k_{B}\ln \mathcal{W}$ based on
classical mechanical models of thermodynamics. The argument is based on the
heat theorem and can be traced back to the second half of the nineteenth
century with the works of Helmholtz and Boltzmann. Despite its simplicity,
this argument has remained almost unknown. We present it in a modern,
self-contained and accessible form. The approach constitutes an important
link between classical mechanics and statistical mechanics.
\end{abstract}

\maketitle

\section{Introduction}

One of the most intriguing equations of modern physics is Boltzmann's
celebrated principle 
\begin{equation}
S_{B}=k_{B}\ln \mathcal{W},  \label{eq:SlogW}
\end{equation}%
where $k_{B}$ is Boltzmann's constant. Despite its unquestionable success in
providing a means to compute the thermodynamic entropy of isolated systems
based on counting the number of available microscopic states $\mathcal{W}$, its
theoretical justification remains obscure and vague in most statistical
mechanics textbooks. In this respect Khinchin commented:\cite{Khinchin} ``All existing
attempts to give a general proof of this postulate must be considered as an
aggregate of logical and mathematical errors superimposed on a general
confusion in the definition of the basic quantities.'' This
lack of a crystal clear proof of Boltzmann's principle puts physics
students, teachers, and in indeed all physicists, in the uncomfortable
position of being forced to accept the formula as a \emph{postulate} that is
necessary to link thermodynamic entropy to microscopic dynamics.

Recent studies in the field of history and foundations of statistical
mechanics \cite{Campisi05} have drawn the attention to the fact that a
similar formula, 
\begin{equation}
S=k_{B}\ln \Phi ,  \label{eq:S=LogPhi-intro}
\end{equation}
emerges naturally from classical mechanics if (a) the ergodic hypothesis is made, (b)
the properties that entropy
should satisfy are appropriately set and the basic quantities are
consistently defined. The quantity $\Phi $ is the volume in phase space
enclosed by a hyper-surface of constant energy $E$.

Equation (\ref{eq:S=LogPhi-intro}) is valid for both small and large systems
and coincides with the Boltzmann formula for large systems. Hence the
derivation of Eq. (\ref{eq:S=LogPhi-intro}) provides the missing proof of
Eq. (\ref{eq:SlogW}). The basic argument underlying the derivation of Eq. (%
\ref{eq:S=LogPhi-intro}) can be traced back to as early as the second half
of the nineteenth century in the works of Helmholtz and Boltzmann.\cite{Helmholtz,Bol84}

The purpose of this article is to provide a widely accessible modern presentation
of the original argument of Helmholtz and Boltzmann \cite{Helmholtz,Bol84} and of its 
recent developments \cite{Campisi05}, that could be used in the
classroom. We derive Boltzmann's principle from classical mechanics with one
simple guiding principle, \emph{viz.}, the heat theorem (see Statement 1, below),
and one central assumption, namely, the ergodic hypothesis.

In Sec. \ref{sec:thermodynamics} we briefly review basics of thermodynamics.
We give concise formulations of the first and second laws of thermodynamics
and introduce the heat theorem. We then construct a one-dimensional
mechanical model of thermodynamics in Sec. \ref{sec:1D} according to the
work of Helmholtz.\cite{Helmholtz} The concepts of ergodicity and
microcanonical probability distribution emerging naturally in this model are
introduced in Sec. \ref{sec:ergos}. We then generalize the one-dimensional
model to more realistic Hamiltonian systems of $N$-particles in
three-dimensions in Sec. \ref{sec:multi}. At this stage the ergodic hypothesis is made and Eq. (\ref{eq:S=LogPhi-intro}) is derived.
In Sec. \ref{sec:adiabatic}, we point out that the mechanical
entropy of Eq. (\ref{eq:S=LogPhi-intro}) does not change during quasi-static
processes in isolated systems, in agreement with the second law of thermodynamics. Non
quasi-static processes that can lead to an increase of entropy have been
treated elsewhere.\cite{CampisiSHPMP2,CampisiPRE08} In Sec. \ref{sec:boltzmann} the Boltzmann principle is derived. A summary and some remarks
concerning the validity of the ergodic hypothesis are given in Sec. \ref{sec:conlcusions}.

In the text we present the line of reasoning and the main results, while
proofs and problems are provided in the appendices.

\section{\label{sec:thermodynamics}Clausius Entropy}

The purpose of this paper is to construct a \emph{classical mechanical} 
analog of \emph{thermodynamic entropy}. 
To this end it is necessary to give a clear account of the definition of entropy
in thermodynamics.  We now review the first
and second laws of thermodynamics in the formulation given by Clausius (see Ref. \cite{Uffink01}). The latter gives the definition of thermodynamic entropy.

\subsection*{First Law of Thermodynamics}

In its differential form the first law of thermodynamics reads: \cite{callen}
\begin{equation}
dE=\delta Q+\delta W,  \label{First}
\end{equation}
where $dE$ is the change in internal energy, $\delta Q$ is the heat added to
the system and $\delta W$ is the work done on the system during an
infinitesimal transformation. The first law is the energy conservation law
applied to a system in which there is an exchange of energy by both work and
heat.

Of crucial importance for the understanding of the first law is that $\delta
Q$ and $\delta W$ are \emph{inexact} differentials, whereas $dE$ is \emph{exact}. The internal energy $E$ is a \emph{state variable}, namely a
quantity that characterizes the thermodynamic equilibrium \emph{state} of
the system.
On the other hand, $W$ and $Q$ are quantities that characterize
thermodynamic energy \emph{transfers} only and are not properties of the
state of the system.\cite{Chandler}

Exactness of differentials is best understood in terms of their integral along a path in the system's state space. The differential is exact if and only if the integral depends only on the end points of the path. Accordingly, the integral of an inexact differential along two different paths with same end points may take on different values. The interested readers find a summary of the formal definition of differential forms and their major properties in Appendix \ref{app:dif-form}.

\subsection*{Second Law of Thermodynamics}

The second law of thermodynamics can be conveniently summarized as three
statements.

\begin{statement}
The differential ${\delta Q}/{T} $ is exact.
\end{statement}

Statement 1, is one
of the most important statements of thermodynamics: Although $\delta Q$ is
not an exact differential, when it is divided by the absolute temperature $T$,
an exact differential is obtained.

This is equivalent to stating that that there exist a state function $S$,
such that 
\begin{equation}
\frac{\delta Q}{T}=dS.  \label{Second}
\end{equation}
The function $S$ is called the \emph{thermodynamic entropy} of the system.

Statement 1 can be expressed in an equivalent way also in \emph{integral form},
by stating that the integral of $\delta Q/T$ along a path connecting a
state $A$ to a state $B$ in the state variables' space, does not depend on
the path but only on its endpoints $A$ and $B$.\cite{note1} This in turn says that there
exists a state function $S$ (i.e. the thermodynamic entropy), such that 
\begin{equation}
\int_{A}^{B}\frac{\delta Q}{T}=S(B)-S(A).  \label{deltaS}
\end{equation}

From Eq. (\ref{First}) it is $\delta Q =dE -\delta W$. In general, the work
is performed by changing a certain number of external parameters $\lambda_i$, e.g. volume, magnetic field, electric field. Then the work $\delta W$ is
given by $-\sum_i F_i d\lambda_i$, where $F_i$ denote the corresponding
conjugate forces, i.e., pressure, magnetization, electric polarization,
respectively. Therefore it is: 
\begin{equation}
\delta Q = dE+\sum_i F_i d\lambda_i
\end{equation}
Without loss of generality, in the following we will restrict ourselves to the case of
only one external parameter $V$ with conjugate force $P$:\cite{note2}
\begin{equation}
\delta Q = dE+ P dV
\end{equation}
In this case Statement 1 can be re-expressed as:
\begin{equation}
(dE + PdV)/T = \emph{exact differential} =dS,  \label{heatthm}
\end{equation}
This is referred to in the literature as the \emph{heat theorem}. \cite{Gallavotti}
The heat theorem can be re-expressed in equivalent terms as: \emph{there exists a function $S(E,V)$ such that:}\cite{note3}
\begin{equation}
\frac{\partial S}{\partial E} = \frac{1}{T}, \qquad \frac{\partial S}{\partial V} =\frac{P}{T}.  \label{heatthmb}
\end{equation}

It is worth emphasizing that any inexact differential, like for example $\delta Q=dE+PdV$, does not enjoy the same property: it is impossible, in general, to find a function of state $Q(E,V)$ such that $\partial Q/ \partial E =1$ and $\partial Q/ \partial V =P$.

The following two statements, regarding the function $S$ complete Clausius's
form of the second law:

\begin{statement}
For a quasi-static process occurring in a thermally isolated system, the entropy
change between two equilibrium states is zero, 
\begin{equation}
\Delta S=0.  \label{delS0}
\end{equation}
\end{statement}

\begin{statement}
For a non quasi-static process occurring in a thermally isolated system, the entropy
change between two equilibrium states is nonnegative, 
\begin{equation}
\Delta S\geq 0.  \label{delS+}
\end{equation}
\end{statement}

A crucial point that must not be overlooked is that Statements 2 and 3
pertain to \emph{thermally isolated} systems. This means that the system is
not in contact with a thermal bath, by means of which one could in principle
control either its temperature or its energy. Thus the processes mentioned
therein are processes in which only the external parameter $V$ is varied in
a controlled way and there is no control over the variable $E$. In Statement
2 the change of the parameter $V$ is so slow that at any instant of time the
system is almost at equilibrium (quasi-static process). In Statement 3, this
requirement is relaxed.

\section{\label{sec:1D}One-dimensional mechanical models of thermodynamics}

In this section we construct a one-dimensional classical mechanical analogue
of Clausius thermodynamic entropy. This construction dates back to Helmholtz
\cite{Campisi05,Helmholtz,Gallavotti} and is based on the heat theorem ($\ref{heatthm}$).

\begin{figure}[tbp]
\begin{center}
\includegraphics[width=8cm]{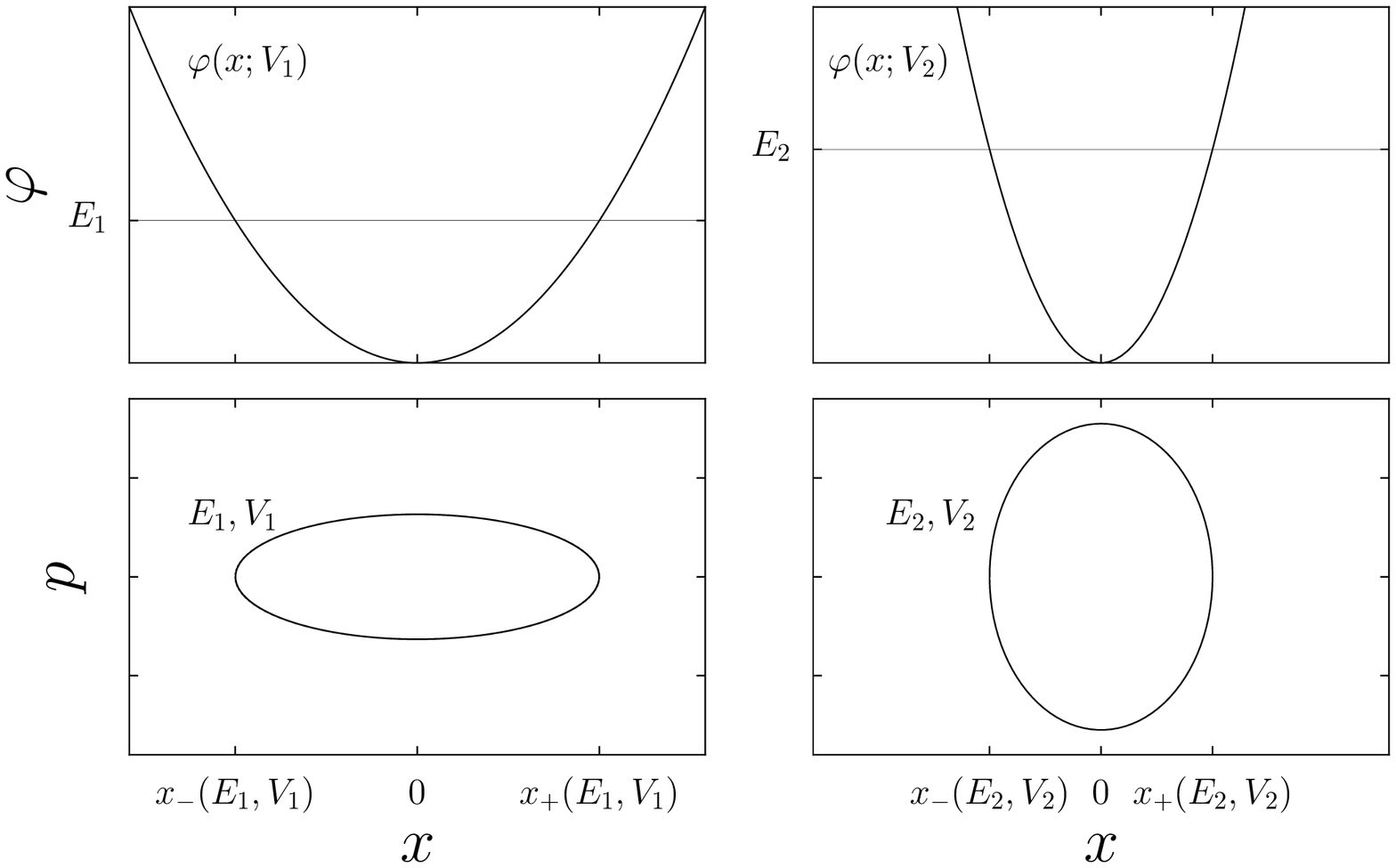}
\end{center}
\caption{Point particle in the $U$-shaped potential $\protect\varphi%
(x;V)=mV^2x^2/2$. Top left panel: shape of the potential for a certain $%
V=V_1 $. Top right panel: shape of the potential for a certain $V=V_2$.
Bottom left panel, phase space orbit corresponding to the potential $\protect%
\varphi(x;V_1)$ at energy $E_1$. Bottom right panel, phase space orbit
corresponding to the potential $\protect\varphi(x;V_2)$ at energy $E_2$. The
two quantities $E,V$, uniquely determine one ``state", i.e., one closed orbit
in the phase space.}
\label{fig:1}
\end{figure}

Consider a point particle of mass $m$ and coordinate $x$ moving in a $U$-shaped potential $\varphi(x)$, as illustrated in Fig. \ref{fig:1}.

In order to allow for the possibility of doing work on the particle by means
of an external intervention, we assume the potential $\varphi$ to depend on
some externally controllable parameter $V$: $\varphi=\varphi(x;V)$. As an
example one could think of a pendulum whose length can be changed at will by
an experimenter while the pendulum oscillates. In this case $V$ would denote
the length of the pendulum. The Hamiltonian of the system is: 
\begin{equation}
H(x,p;V)=K(p)+\varphi(x;V),  \label{ham}
\end{equation}
where $K(p)=p^{2}/2m$ is the kinetic energy and $p$ is the momentum.

Now that our mechanical system is defined, we have to specify its ``internal
energy'' $E$, ``temperature'' $T$, and the ``force'' $P$ conjugate to the
external parameter $V$.

For the internal energy we simply take the energy $E$ given by the
Hamiltonian. For a fixed $V$ the particle's energy $E$ is a constant of
motion. For the sake of simplicity we chose the gauge of the potential in
such a way that the minimum of the potential is $0$ regardless of the value
of $V$.

Once $V$ and $E$ are specified, the orbit in phase space of the particle is
fully determined: we say that $E$ and $V$ are the system's state variables. \cite{note4}
The particle moves back and forth between the two turning points $%
x_{\pm}(E,V)$, and draws closed orbits in phase space with a certain period $%
\tau(E,V)$. See Fig. \ref{fig:1}.

Now that the state variables are fixed we have to define the corresponding
temperature and conjugate force. In agreement with the common understanding
of temperature as a measure of the speed of the particles, we take the
temperature to be proportional to the kinetic energy averaged over one
period: 
\begin{equation}
T(E,V):=\frac{2}{k_{B}\tau (E,V)}\int_{0}^{\tau (E,V)}dtK(p(t;E,V))
\label{eq:kin1}
\end{equation}%
Thanks to the the factor $1/k_{B}$, $T$ has the correct dimensions of a
temperature. For the conjugate force we take the time average of $-\frac{%
\partial \varphi }{\partial V}$ \cite{Landau1}, i.e., 
\begin{equation}
P(E,V):=-\frac{1}{\tau (E,V)}\int_{0}^{\tau (E,V)}dt\frac{\partial \varphi
(x(t;E,V);V)}{\partial V}  \label{eq:press1}
\end{equation}%

In Eqs. (\ref{eq:kin1},\ref{eq:press1}), $x(t;E,V)$ and $p(t;E,V)$ are the
solution of Hamilton's equations of motion with a fixed $V$, and an
arbitrary initial condition $x_0,p_0$ such that $H(x_0,p_0;V)=E$.

Having identified the mechanical analogues of internal energy, external
parameter, temperature and conjugate force with the quantities $E,V,T,P$,
respectively, we can now ask whether a mechanical analogue of entropy, $S$,
exists. In order to answer this question we must ask, in agreement with
Statement 1 as expressed in Eq. (\ref{heatthmb}), whether there exists a
function $S(E,V)$ such that: 
\begin{equation}
\frac{\partial S}{\partial E}(E,V) = \frac{1}{T(E,V)}, \quad \frac{\partial S%
}{\partial V}(E,V) =\frac{P(E,V)}{T(E,V)}.  \label{eq:HeatTheoMath}
\end{equation}
%---------------------------------------------------------------------------------------------------
%-----------------------------------------------------------------------------------------------------
The answer to this question is given by the following theorem.

\begin{theo}[Helmholtz]
A function $S(E,V)$ satisfying Eq. (\ref{eq:HeatTheoMath}) exists and is
given by 
\begin{equation}
S(E,V)= k_B\log 2\int_{x_-(E,V)}^{x_+(E,V)} dx \sqrt{2m(E-\varphi(x;V))}
\label{eq:S1a}
\end{equation}
\end{theo}

The proof of the theorem is reported in Appendix \ref{app:theo1}, see also Ref.
\cite{Gallavotti} (pp 45--46).

The quantity $\sqrt{2m(E-\varphi(x;V))}$ represents the momentum of the
particle when it is located at $x$. Solving the equation $E=p^2/2m +
\varphi(x;V)$, and taking the positive root, we obtain: 
\begin{equation}
p(x;E,V)= \sqrt{2m(E-\varphi(x;V))}
\end{equation}
Thus the entropy can be rewritten compactly as: 
\begin{equation}
S(E,V)= k_B \log \oint pdx  \label{eq:S1b}
\end{equation}
The integral $\oint pdx$ is called the \emph{reduced action}.\cite{Landau1} It represents
the area $\Phi$ enclosed by the orbit of energy $E$, and parameter $V$ in phase
space: 
\begin{equation}
S(E,V)= k_B \log \Phi(E,V)  \label{eq:S1c}
\end{equation}
where 
\begin{equation}
\Phi(E,V)= \int_{H(x,p;V)\leq E} dp dx.  \label{eq:A1}
\end{equation}
Theorem 1 says that there exists a mechanical counterpart of entropy, and
that this is the logarithm of the phase space volume \emph{enclosed} by the
curve of constant energy $H(x,p;V)= E$.

The fact that there exists a function $S$ satisfying Eq. (\ref{eq:HeatTheoMath})
is a highly non trivial result which tells us that we have
a consistent one-dimensional mechanical model of thermodynamics. Although
this model nicely suggests the deep connection between classical mechanics
and thermodynamic entropy, it is definitely too stylized to model a real
thermodynamic system composed of as many as $10^{23}$ particles. It is
necessary to generalize Helmholtz Theorem to multidimensional systems.

\section{\label{sec:ergos}Ergodicity and Microcanonical Ensemble}

The main ingredients needed for the generalization of our model to more
degrees of freedom are \emph{ergodicity} and the \emph{microcanonical}
ensemble.

We use the one-dimensional example of the previous section to introduce
these important concepts. Imagine we want to calculate the time average of a
phase function $f(x,p)$ over the orbit specified by $E$ and $V$: 
\begin{equation}
\langle f\rangle _{t}:=\frac{1}{\tau }\int_{0}^{\tau }dtf(x(t),p(t)).
\end{equation}%
For simplicity of notation we have dropped the explicit dependence of $\tau $%
, $x(t)$, $p(t)$ and $\langle f\rangle _{t}$ on $E$, $V$.

Since $p=mv=mdx/dt$ the differential $dt$ is: 
\begin{equation}
dt=m\frac{dx}{p(x)}.
\end{equation}%
With this we obtain: 
\begin{equation}
\langle f\rangle _{t}=\frac{2m}{\tau }\int_{x_{-}}^{x_{+}}\frac{dx}{p(x)}%
f(x,p(x))  \label{eq:<f>t_1D}
\end{equation}%
where the factor 2 stems from the fact that the particle goes from $x_{-}$
to $x_{+}$, in a half period, i.e., $\tau /2$. Now consider the following
integral 
\begin{equation}
\int dp\delta \left( p^{2}/2m+\varphi (x;V)-E\right) 
\end{equation}%
where $\delta $ denotes Dirac's delta function. Using the formula $\delta
(f(p))=\sum_{i}\delta (p-p_{i})/|f^{\prime }(p_{i})|$, where the $p_{i}$'s
are the zeroes of $f(p)$, and $\int dp\delta (p-p_{i})$=1, we get: 
\begin{equation}
\int dp\delta \left( p^{2}/2m+\varphi (x;V)-E\right) =2m/p(x)
\end{equation}%
Then Eq. (\ref{eq:<f>t_1D}) becomes: 
\begin{equation}
\langle f\rangle _{t}=\frac{1}{\tau }\int dx\int dp\delta \left(
p^{2}/2m+\varphi (x;V)-E\right) f(p,x)
\label{eq:<f>t_1D_b}
\end{equation}%
where the integration extremes $x_{\pm }$ need not be specified, being
implied by the Dirac $\delta $. The period $\tau $ is given by: 
\begin{eqnarray}
\tau  &=&\int_{0}^{\tau }dt=2\int_{x_{-}}^{x_{+}}\frac{dx}{p(x)}  \notag \\
&=&\int dx\int dp\delta \left( p^{2}/2m+\varphi (x;V)-E\right) 
\label{eq:tau}
\end{eqnarray}%
Hence we arrive at: 
\begin{equation}
\langle f\rangle _{t}=\int dx\int dp\rho (x,p;E,V)f(p,x)
\label{eq:<f>t_1D_c}
\end{equation}%
where we have introduced the phase space probability density function 
\begin{equation}
\rho (x,p;E,V)=\frac{1}{\tau (E,V)}\delta \left( p^{2}/2m+\varphi
(x;V)-E\right) 
\label{eq:rho1d}
\end{equation}%
From Eq. (\ref{eq:rho1d}) it is clear that $\tau $ is the normalization. This $%
\rho (x,p;E,V)$ is called the microcanonical distribution. Eq. (\ref{eq:<f>t_1D_c})
says that the time average of a phase space quantity $f(x,p)$
over one period, is equal to its microcanonical average. This property is
called ergodicity. All one-dimensional systems with a $U$-shaped potential
are ergodic.

\section{\label{sec:multi}Multi particle mechanical models of thermodynamics}

We now can extend the previous treatment to systems of $N$-particles in
three dimensions. The Hamiltonian for an interacting system of $N$-particles
of mass $m$ is 
\begin{equation}
H_N(\mathbf{q},\mathbf{p};V)=K_N(\mathbf{p})+\varphi_N(\mathbf{q};V),
\label{Ham}
\end{equation}
where $K_N(\mathbf{p})= \sum_{i=1}^{3N} p_i^2/2m$ is the kinetic energy, $%
\varphi_N$ is the potential energy, and the coordinates $\mathbf{q}%
=\{q_i\}_{i=1} ^{3N}$ and their conjugate canonical momenta $\mathbf{p}=\{
p_i \}_{i=1}^{3N}$ are $3N$-dimensional vectors.

In analogy with one-dimensional systems with a $U$-shaped potential, we
define the microcanonical probability distribution as 
\begin{equation}
\rho _{N}(\mathbf{q},\mathbf{p};E,V)=\frac{1}{\Omega _{N}(E,V)}\delta \left(
E-H_{N}(\mathbf{q},\mathbf{p};V)\right) ,
\end{equation}%
where $\Omega _{N}(E,V)$ is the normalization: 
\begin{equation}
\Omega _{N}(E,V)=\idotsint d\mathbf{q}d\mathbf{p}\delta \left( E-H_{N}(\mathbf{q},%
\mathbf{p};V)\right) .  \label{eq:Omega}
\end{equation}

Continuing the analogy with one-dimensional systems, we make the following crucial
assumption:

\begin{hypo}
For given $E$ and $V$, the time average $\langle f\rangle _{t}$ 
of any
function $f(\mathbf{q},\mathbf{p})$ is uniquely determined and is equal to its microcanonical average $\langle
f\rangle _{\mu }$, i.e.:
\begin{equation}
\langle f\rangle _{t}=\idotsint d\mathbf{q}d\mathbf{
p}\text{ }\rho_N (\mathbf{q},\mathbf{p};E,V)f(\mathbf{q},\mathbf{p})\doteq \langle f\rangle _{\mu }.
\label{ergN}
\end{equation}
\end{hypo}

In analogy with Eqs.(\ref{eq:kin1},\ref{eq:press1}) we define the
temperature as: 
\begin{equation}
T_{N}(E,V):=\frac{2}{3Nk_{B}}\langle K_{N}\rangle _{t}  \label{eq:kin3N}
\end{equation}%
and the conjugate force as 
\begin{equation}
P_{N}(E,V):=-\left\langle \frac{\partial \varphi _{N}}{\partial V}%
\right\rangle _{t}.  \label{eq:press3N}
\end{equation}%
We ask, in agreement with Statement 1 as expressed in Eq. 
(\ref{heatthmb}), whether there exists a function $S_{N}(E,V)$ such that 
\begin{equation}
\frac{\partial S_{N}}{\partial E}(E,V)=\frac{1}{T_{N}(E,V)},\quad \frac{%
\partial S_{N}}{\partial V}(E,V)=\frac{P_N(E,V)}{T_{N}(E,V)}.
\label{eq:HeatTheoMath2}
\end{equation}
The answer is given by the following theorem:

\begin{theo}[Helmholtz, Generalized]
A function $S_N(E,V)$ satisfying Eq. (\ref{eq:HeatTheoMath2}) exists and is
given by: 
\begin{equation}
S_N(E,V)= k_B\log \Phi_N(E,V)  \label{eq:S3Na}
\end{equation}
where 
\begin{equation}
\Phi_N(E,V):=\idotsint_{H_N(\mathbf{q},\mathbf{p}) \leq E} d\mathbf{q} d\mathbf{p}
\label{eq:Phi}
\end{equation}
\end{theo}

The proof, which is based on the equipartition theorem, is given in Appendix %
\ref{app:theo2}. See also Ref. \cite{Campisi05}.

We draw the attention to the fact that, unlike temperature and conjugate
force, $S_N$ is not in the form of the time average of some phase function $%
f(\mathbf{q},\mathbf{p})$

Theorem 2 says that ergodic systems constitute ideal mechanical models of
thermodynamics. One can define their state variables by $E$ and $V$ as in
thermodynamics. Moreover, one can define their temperature and conjugate
force straightforwardly as functions of the state variables. Surprisingly,
the heat differential $(dE+P_NdV)/T_{N}$ is exact, allowing for a
consistent and logical definition of entropy $S_{N}$.

\section{\label{sec:adiabatic}Adiabatic Invariance}
According to Theorem 2, $S_N$ complies with the first law of thermodynamics and statement 1 of the
second law of thermodynamics: Is the construction
consistent with statements 2 and 3 of the second law of thermodynamics as well?

Let us focus on statement 2. In order for $S_{N}$ to be consistent with this
statement it is necessary that if the parameter $V$ is varied very slowly in
time (much slower than any time scale of the system dynamics) from a certain 
$V_{0}=V(t_{0})$ to a certain $V_{f}=V(t_{f})$, the corresponding change of
the entropy $S_{N}$ is null. Note that, by allowing for a time dependence of 
$V$, the system's Hamiltonian now becomes  time dependent, and energy is not
conserved. Let the system be at $t=t_{0}$, in $\mathbf{q}_{0},\mathbf{p}_{0}$%
. Under the time dependent Hamiltonian 
\begin{equation}
H_{N}(\mathbf{q},\mathbf{p};V(t))=K_{N}(\mathbf{p})+\varphi _{N}(\mathbf{q}%
;V(t)),  \label{Ham_t}
\end{equation}%
it evolves to a new phase space point $\mathbf{q}_{f}(\mathbf{q}_{0},\mathbf{%
p}_{0}),\mathbf{p}_{f}(\mathbf{q}_{0},\mathbf{p}_{0})$, where we made
explicit the dependence on the initial condition of the evolved phase space
point. Then the energy at time $t_{f}$ is $E_{f}=H_{N}(\mathbf{q}_{f}(%
\mathbf{q}_{0},\mathbf{p}_{0}),\mathbf{p}_{f}(\mathbf{q}_{0},\mathbf{p}%
_{0});V(t_{f}))$. It is known \cite{Berdichevsky} that, for \emph{ergodic
systems}, the energy reached at the end of a very slow protocol depends only
on the initial energy $E_{0}=H_{N}(\mathbf{q}_{0},\mathbf{p}_{0};V(t_{0}))$,
and is determined by solving the following equation for $E_{f}$: 
\begin{equation}
\Phi _{N}(E_{0},V_{0})=\Phi _{N}(E_{f},V_{f})
\end{equation}%
That is, the quantity $\Phi _{N}(E,V)$ does not change in the course of
time, when $V$ is varied infinitely slowly. This property is called in
classical mechanics \emph{adiabatic invariance}. Since it is $%
S_{N}(E,V)=k_{B}\log \Phi _{N}(E,V)$, and $\Phi _{N}(E,V)$ is an adiabatic
invariant, it is evident that $S_{N}$ is an adiabatic invariant too. Namely,
it does not change if $V$ is changed very slowly in time. Thus $S_{N}$
complies with Statement 2. For completeness in Appendix \ref{app:AdInv} we
provide a proof of adiabatic invariance of $\Phi _{N}$. See also Ref. \cite{Berdichevsky} (pp. 27--30).

It is also possible to prove that, in an averaged sense, $S_{N}$ complies
with Statement 3, as well.\cite{CampisiSHPMP2,CampisiPRE08} In this case
one has to consider the average change of entropy, because, for fast
transformations, the final energy is not uniquely determined by the initial
energy, and depending on the initial conditions, one ends up with different
final energies, i.e., different final entropies.

\section{\label{sec:boltzmann}Boltzmann Principle}

For a system composed of a very large number $N$ of particles which interact
via short range forces, the phase space volume $\Phi_N(E)$ approaches an exponential behavior $\Phi_{N}(E) \propto e^E$. Since $\Omega_N=\partial \Phi_N /\partial E$ (see Eq. (\ref{eq:Omega})), it is $\Phi_{N} \propto \Omega_{N}$ (see Ref. \cite{Huang}, p.148).

The quantity $\Omega _{N}$, defined in Eq. (\ref{eq:Omega}), represents the
measure of the shell of constant energy $H_{N}(\mathbf{q},\mathbf{p};V)=E$.
As such it is proportional the number $\mathcal{W}$ of micro states
compatible with the given energy $E$. (According to semiclassical theory each micro state occupies a volume $h^{3N}$ of phase space, where $h$ is Planck's constant.\cite{Landau5} By introducing an arbitrary energy scale $\Delta E$, the number $\mathcal{W}$ is given by $\mathcal{W} = \Omega_N \Delta E /h^{3N}$). Thus, for very large $N$, 
\begin{equation}
\Phi_{N} \propto \Omega_{N} \propto \mathcal{W},\quad N\gg 1
\end{equation}
By taking the logarithm, we have
\begin{equation}
S_{N} \simeq k_{B}\ln \mathcal{W}=S_B,\quad N\gg 1
\label{eq:S_N=lnW}
\end{equation}
except for an irrelevant constant. Eq. (\ref{eq:S_N=lnW}) says that for large ergodic systems composed of particles interacting via short range forces, the differential of Boltzmann entropy is equal to the differential $\delta Q/T$. Hence it can be identified with Clausius entropy. This is a proof of Boltzmann principle.

\section{\label{sec:conlcusions}Conclusions}

Given an ergodic system, it is possible to specify its thermodynamic state by
means of the total energy $E$ and the external parameter $V$. Given the
state $E,V$, we can unambiguously define the quantities $T_N(E,V)$ and $%
P_N(E,V)$. Once these are identified with the system temperature and
conjugate force, one can ask whether, as prescribed by the heat theorem, the
combination: 
\begin{equation}
\frac{dE+P_NdV}{T_N}  \notag
\end{equation}
is an exact differential. Surprisingly the answer is positive, meaning that
there exist a function $S_N(E,V)$ which can be identified with the
thermodynamic entropy of the system. The generalized Helmholtz theorem says
that this is given by the logarithm of the volume $\Phi_N(E,V)$ of phase
space \emph{enclosed} by the hyper-surface of energy $H(\mathbf{q},\mathbf{p}%
;V)=E$. For macroscopic systems this entropy coincides with Boltzmann
entropy, thus revealing the rationale of Boltzmann principle.

The entropy in Eq. (\ref{eq:S3Na}) is sometimes referred to in the
literature as Hertz entropy.\cite{Dunkel06,adib04} Hertz \cite{Hertz1,Hertz2} 
derived it from the requirement of adiabatic invariance
(see also Refs. \cite{Berdichevsky,Munster,Rugh01}), whereas we have
derived it here from the heat theorem. Its fundamental character is also recognized
in Ref. \cite{Schluter48} where its property of being a \emph{canonical invariant} is emphasized and in 
Ref. \cite{Pearson85} which highlights its compliance with the equipartition theorem, and the fact that it is a positive and increasing function of the energy\cite{TalknerHanggiMorillo07,noteTemp}.
The entropy in Eq. (\ref{eq:S3Na}) also appears in Gibbs seminal book.\cite{Gibbs02}
However its connection with the heat theorem has not previously recognized.

The most crucial point of the derivation of Boltzmann principle is the introduction of the ergodic
hypothesis. 
Although this hypothesis is generally believed to be true for
real macroscopic systems, its mathematical proof is a formidable challenge which has
been achieved only in few special cases.\cite{Lebowitz73} A proof that a gas of elastically colliding hard spheres is ergodic was announced in 1963 by Sinai. \cite{Sinai63} However the full proof was not published and the problem is still open (see Ref. \cite{UffinkCompendium} for a more detailed discussion). Nonetheless ergodicity of hard spheres systems seems plausible as indicated also by recent numerical simulations (see Sec. IV of Ref. \cite{CampisiPRE09}). The hypothesis cannot be true in the case of real crystals. Here the nuclei remain close to their lattice site, preventing them from sampling the whole energy hypersurface homogeneously (this being a necessary condition for ergodicity).

In regard to these difficulties, it is worth pointing out that  the present derivation of Boltzmann principle does not make use of the fact that the average of \emph{any} arbitrary phase function be equal to its microcanonical average, as required by the ergodic hypothesis. It \emph{only} requires that the time average of $K$ and $-\partial \varphi/\partial V$, be equal to their microcanonical averages (see the proof of Theorem 2 in appendix \ref{app:theo2}). Thus the ergodic hypothesis can be greatly relaxed by requiring the much less stringent condition that $T_N$ and $P_N$, Eqs. (\ref{eq:kin3N},\ref{eq:press3N}), can be calculated as microcanonical averages.\cite{note5}
In this case, the Clausius entropy can still be calculated via the Hertz formula (\ref{eq:S3Na}).

\section*{Acknowledgements}

We wish thank the Texas Section of the American Physical Society for the
``Robert S. Hyer Recognition for Exceptional Research'' presented at its
Fall 2008 meeting, and Cosimo Gorini for reading the manuscript. We also would like to thank Prof. Randall B. Shirts for providing the translation of Ref. \cite{Schluter48}. Valuable remarks from the anonymous referees are gratefully acknowledged.

\appendix

\section{\label{app:dif-form}Differential forms: Brief review of definitions
and main results}

A differential form $\omega$ in a connected subset $\mathcal{A}$ of $\mathbb{%
R}^2$ is formally written as: 
\begin{equation}
\omega = \mathcal{M}(x_1,x_2)dx_1 + \mathcal{N}(x_1,x_2)dx_2
\end{equation}
where $\mathcal{M}(x_1,x_2), \mathcal{N}(x_1,x_2)$ are two functions on $\mathcal{A}$ and $(x_1,x_2)$ are
the coordinates in $\mathbb{R}^2$. \cite{note6}

Given a curve $\psi:[s_0,s_1] \rightarrow \mathcal{A}$, 
\begin{equation}
\psi(s)=(\psi_1(s),\psi_2(s))
\end{equation}
the integral of $\omega$ along the curve $\psi$, is defined as: 
\begin{equation}
\int_\psi \omega = \int_{s_0}^{s_1} [\mathcal{M}(\psi(t))\psi^{\prime
}_1(t)+\mathcal{N}(\psi(t))\psi^{\prime }_2(t)]dt
\end{equation}
where $\psi^{\prime }_{1,2}$ are the derivatives of $\psi_{1,2}$.

A differential form $\omega$ is said to be \emph{exact}, if there exist a
function $G: \mathcal{A} \rightarrow \mathbb{R}$ such that: 
\begin{equation}
\omega = dG
\end{equation}
that is: 
\begin{equation}
\frac{\partial G}{\partial x_1}(x_1,x_2) = \mathcal{M}(x_1,x_2),\quad \frac{\partial G}{\partial
x_2}(x_1,x_2) = \mathcal{N}(x_1,x_2)
\end{equation}
$G$ is called a \emph{primitive} for the differential form. \newline

Let $\Sigma(A,B)$ be the set of all curves connecting
the point $A \equiv (a_{1},a_{2})$ to the point $B \equiv (b_{1},b_{2})$ in $%
\mathcal{A}$. A differential form is exact if and only if for any couple of
points $A$ and $B$ in $\mathcal{A}$ and curves $\psi$
and $\phi$ in $\Sigma(A,B)$, it is 
\begin{equation}
\int_\psi \omega =\int_\phi \omega
\end{equation}

The integral of an exact differential form along any curve $\psi$ connecting 
$A$ to $B$ does not depend on the curve $\psi$, but
only on the ending points, and is given by: 
\begin{equation}
\int_\psi \omega = \int_\psi dG = G(B)-G(A).
\end{equation}

The following statement also holds: A differential form is exact if and only
if its integral along any closed curve is null.

If the functions $\mathcal{M}$ and $\mathcal{N}$ are of class $C^1$ (i.e. they are
differentiable), then a necessary condition for the form $\omega$ to be
exact is that: 
\begin{equation}
\frac{\partial \mathcal{M}}{\partial x_2} = \frac{\partial \mathcal{N} }{\partial x_1}
\label{eq:Schwarz}
\end{equation}
In this case the differential form $\omega$ is said to be \emph{closed}.

\section{\label{app:theo1}Proof of Theorem 1}

It is known \cite{Landau1} that in one-dimensional systems confined in a $U$%
-shaped potential, the period $\tau $ of the orbit is equal to the
derivative of the area $\Phi$ in phase space enclosed by the orbit with respect
to energy. 
\begin{equation}
\tau =\frac{\partial \Phi}{\partial E}  \label{eq:tau=dA/dE}
\end{equation}%
One simple way to prove this relation is by expressing the area as $%
\Phi(E,V)=\int dpdx\theta (E-H(x,p;V))$, where $\theta (x)$ is Heaviside step
function [$\theta (x)=1$ if $x\geq 0$, $\theta (x)=0$ if $x<0$]. Taking the
derivative with respect to $E$, and using the relation $\delta (x)=d\theta
(x)/dx$, gives $\tau $ (see Eq. (\ref{eq:tau})). Using Eq. (\ref%
{eq:tau=dA/dE}) and Eq. (\ref{eq:S1c}) we obtain: 
\begin{equation}
\frac{\partial S}{\partial E}=k_{B}\frac{\tau }{\Phi}.  \label{eq:dS/dE=tau/A}
\end{equation}%
From Eq. (\ref{eq:<f>t_1D}), we obtain the relation: 
\begin{equation}
2\langle K\rangle _{t}=\frac{\Phi}{\tau }  \label{eq:equiPart1D}
\end{equation}%
from which, using Eq. (\ref{eq:dS/dE=tau/A}), we get: 
\begin{equation}
\frac{\partial S}{\partial E}=\frac{k_{B}}{2\langle K\rangle _{t}}
\end{equation}%
Similarly we also get: 
\begin{equation}
\frac{\partial S}{\partial V}=-\frac{k_{B}}{2\langle K\rangle _{t}}%
\left\langle \frac{\partial \varphi }{\partial V}\right\rangle _{t}
\end{equation}%
Using Eqs. (\ref{eq:kin1},\ref{eq:press1}) we obtain: 
\begin{eqnarray}
\frac{\partial S}{\partial E} &=&\frac{1}{T} \\
\frac{\partial S}{\partial V} &=&\frac{P}{T}
\end{eqnarray}%

\section{\label{app:theo2}Proof of Theorem 2}

The proof of Theorem 2 makes use of the multidimensional version of Eq. (\ref%
{eq:tau=dA/dE}), that is: 
\begin{equation}
\Omega_N =\frac{\partial \Phi_N}{\partial E}  \label{eq:Omega=dPhi/dE}
\end{equation}
This can be proved, in a similar way, by expressing $\Phi_N$ as $\int d%
\mathbf{q}d\mathbf{p} \theta(E-H(\mathbf{q},\mathbf{p};V))$ and using
the relation $\delta(x)=d \theta(x)/dx$. The equipartition theorem \cite{Huang}
\begin{equation}
\frac{2\left \langle K \right\rangle_\mu}{3N} = \frac{\Phi_N}{\Omega_N}
\label{eq:equiPart}
\end{equation}
is the generalization of Eq. (\ref{eq:equiPart1D}) to many dimensions. Using
(\ref{eq:equiPart}) and (\ref{eq:Omega=dPhi/dE}) with Eq. (\ref{eq:S3Na}) we
get: 
\begin{equation}
\frac{\partial S_N}{\partial E} = \frac{3Nk_B}{2 \langle K_N \rangle_\mu}
\end{equation}
In a similar way we also get: 
\begin{equation}
\frac{\partial S_N}{\partial V} = -\frac{3Nk_B}{2 \langle K_N \rangle_\mu}
\left \langle \frac{\partial \varphi_N}{\partial V} \right \rangle_\mu
\end{equation}

Using Eqs. (\ref{eq:kin3N},\ref{eq:press3N}) with the ergodic hypothesis, we finally arrive at: 
\begin{eqnarray}
\frac{\partial S_N}{\partial E}&=& \frac{1}{T_N} \\
\frac{\partial S_N}{\partial V}&=& \frac{P_N}{T_N}
\end{eqnarray}

\section{\label{app:AdInv}Proof of adiabatic invariance of $\Phi_N$}

We consider the time-dependent Hamiltonian 
\begin{equation}
H_{N}(\mathbf{q},\mathbf{p};V(t))=K(\mathbf{p})+\varphi (\mathbf{q};V(t)).
\label{HNt}
\end{equation}%
To prove that $\Phi _{N}$ in Eq. (\ref{eq:Phi}) is an adiabatic invariant we
first take the total time derivative of the Hamiltonian $H_N(\mathbf{q,p};V(t))
$ in Eq. (\ref{HNt}), 
\begin{equation}
\frac{dH_{N}(\mathbf{q,p};V)}{dt}=\frac{\partial H_{N}(\mathbf{q,p};V)}{%
\partial V}\frac{dV}{dt}
\label{eq:dH/dt}
\end{equation}%
where the terms involving $\mathbf{\dot{q}}$ and $\mathbf{\dot{p}}$ cancel
by Hamilton's equations.\cite{goldstein} The derivative ${dV}/{dt}$
changes slowly in time, but ${dH_{N}}/{dt}$ and ${\partial H_{N}}/{\partial V}$
can change rapidly because of their dependence on $\mathbf{q}(t)
$ and $\mathbf{p}(t)$. To eliminate the fast variables $\mathbf{q,p}$ we
take the average of Eq. (\ref{eq:dH/dt}) with respect to the microcanonical
ensemble, which gives%
\begin{equation}
\left\langle \frac{dH_{N}}{dt}\right\rangle _{\mu }=\left\langle \frac{%
\partial H_{N}}{\partial V}\right\rangle _{\mu }\frac{dV}{dt}.  \label{dE}
\end{equation}%
By Liouville's theorem \cite{goldstein} the average on the left-hand side of
Eq. (\ref{dE}) is 
\begin{equation}
\left\langle \frac{dH_{N}}{dt}\right\rangle _{\mu }=\frac{dE}{dt}.
\label{dEdt}
\end{equation}%
The microcanonical average in Eq. (\ref{ergN}) on the right-hand side of Eq.
(\ref{dE}) is 
\begin{eqnarray}
\left\langle \frac{dH_{N}}{dV}\right\rangle _{\mu } &=&
\idotsint d\mathbf{q}d\mathbf{p}\text{ }\rho_N (\mathbf{q},\mathbf{p},E,V)%
\frac{\partial H_N(\mathbf{q},\mathbf{p},V)}{\partial V}  \notag \\
&=&-\frac{1}{\Omega _{N}}\frac{\partial \Phi _{N}}{\partial V},  \label{dHdV}
\end{eqnarray}%
where $\Phi _{N}$ and $\Omega _{N}$ are given in Eqs. (\ref{eq:Omega}) and (\ref{eq:Phi}), respectively. Substituting Eqs. (\ref{dEdt})\ and (\ref{dHdV})
into Eq. (\ref{dE}) and using $\Omega _{N}=\partial \Phi _{N}/\partial E$,
we obtain%
\begin{equation}
\frac{d\Phi _{N}}{dt}\equiv \frac{\partial \Phi _{N}}{\partial E}\frac{dE}{dt%
}+\frac{\partial \Phi _{N}}{\partial V}\frac{dV}{dt}=0,  \label{dPhidt}
\end{equation}%
which shows that $\Phi _{N}$ is constant and therefore an adiabatic
invariant.

\section{\label{app:probs}Problems}

Consider the following Hamiltonian of a one dimensional harmonic oscillator
with angular frequency $V$ (see Fig. \ref{fig:1}) 
\begin{equation}
H(x,p;V)= \frac{p^2}{2m}+\frac{mV^2x^2}{2}
\end{equation}
(a) Calculate the area $\Phi(E,V)$ enclosed by the trajectory of energy $E$ and
angular frequency $V$. Using Eq. (\ref{eq:tau=dA/dE}) check that the period
of the orbit is, as expected, given by $\tau(E,V)= 2\pi/ V$.\newline
(b) Using Eqs. (\ref{eq:kin1}, \ref{eq:press1}) show that $k_B T(E,V)=E$, $%
P(E,V)=-E/V$\newline
(c) Show that the differential form $dE+P dV$, with $P(E,V)$ as in (b) is
not exact. (Hint: show that Eq. (\ref{eq:Schwarz}) is not satisfied.) Show
that the integral of $dE+P dV$ over the rectangular path with corners $%
(E_0,V_0),(E_0,V_1),(E_1,V_1),(E_1,V_0)$, and $E_0 \neq E_1$ $V_0 \neq V_1$,
is not null.\newline
(d) Consider the differential form $\omega=(1/T)dE + (P/T) dV$, with $%
P(E,V),T(E,V)$ as in (b). Find a primitive function $S(E,V)$ for $\omega$.
Show that, apart from an additive constant, it is $S(E,V)=\log \Phi(E,V)$, as
dictated by Theorem 1. Check that Eq. (\ref{eq:Schwarz}) is satisfied.

%\bibliographystyle{amjpMichele}
%\bibliography{thebibliography}

\begin{thebibliography}{10}
\bibitem{note1}
A path in the state variables' space corresponds to a transformation that leads the system from $A$ to $B$ through a sequence of almost equilibrium states. Accordingly, the change $\delta Q$ has to be understood as a quasi-static change.
\bibitem{note2}
For the common case of work due to expansion and compression the external parameter is the volume and its conjugate force is the pressure (for this reason the notation $V$ $P$ is adopted). In general $V$ and $P$ stand for any conjugate pair of ``displacement'' and ``force'' depending on the specific nature of the work done on the system (e.g., magnetic field and magnetization).
\bibitem{note3}
To use the language of vector fields, this says that the field 
$\vec{\mathcal F}:= (1/T,F/T)$
is a \emph{conservative} vector field. That is, there exists a \emph{potential} function $S(E,V)$ such that
$\vec{\mathcal F}=\vec{\nabla} S$
where $\vec{\nabla} $ is the gradient operator in the space $(E,V)$. For this reason
the entropy can be understood as a \emph{thermodynamic potential}.
\bibitem{note4}
The variables $(E,V)$ should be understood as the system's ``thermodynamic'' state variables, and should not be confused with the system's ``mechanical'' state variables $(p,x)$.
\bibitem{noteTemp}
The conditions $\Phi \geq 0, \partial_E \Phi = \Omega \geq 0$ ensure that the temperature derived from the Hertz entropy, i.e. $T= (\partial_E \log \Phi)^{-1}=\Phi/\Omega$ be definite positive.
\bibitem{note5}
The simplest example of a non-ergodic system that satisfies this weaker condition is that of a 1-d particle in a symmetric double well potential. For energies below a certain critical value $E_c$ the curve of constant energy splits into two disjoint curves $\gamma_l$ $\gamma_r$, with the phase space trajectory covering only one of them (for energy above $E_c$, there is only one trajectory and ergodicity holds). Since $\gamma_l$ and $\gamma_r$ are one the mirror image of the other, the time averages of even functions of $x$, such as $K$ and $-\partial \varphi / \partial V$ are not affected by which of the two curves the motion actually takes place on. It follows then, that such averages can be calculated as microcanonical averages (that is over the curve $\gamma_l \cup \gamma_r$).
\bibitem{note6}
In the main text the role of $(x_1,x_2)$ is played by the state variables $(E,V)$.
\end{thebibliography}

\end{document}